\documentclass[prd,showpacs,superscriptaddress,twocolumn]{revtex4}
\usepackage{amsfonts}
\usepackage{amsbsy}
\usepackage{mathrsfs}
\usepackage{graphicx}
\usepackage{amsmath}
\def\be{\begin{equation}}
\def\ee{\end{equation}}
\def\bea{\begin{eqnarray}}
\def\eea{\end{eqnarray}}

\def\ba{\begin{array}}
\def\ea{\end{array}}

\def\lsim{\mathrel{\rlap{
\lower4pt\hbox{\hskip-3pt$\sim$}}
\raise1pt\hbox{$<$}}}     
\def\gsim{\mathrel{\rlap{
\lower4pt\hbox{\hskip-3pt$\sim$}}
\raise1pt\hbox{$>$}}}            
\def\vec#1{\mbox{\boldmath $#1$}}

\begin{document}
\title{Transverse $\Lambda^0$ polarization in inclusive photoproduction:
\\ quark recombination model}
\author{I. Alikhanov}
\thanks{ialspbu@mail.ru}
\affiliation{Saint Petersburg State University, Saint Petersburg, 198904, Russia}
\author{O. Grebenyuk}
\thanks{olegreb@pcfarm.pnpi.spb.ru}
\affiliation{Petersburg
Nuclear Physics Institute, Gatchina, Leningrad District, 188350,
Russia}
\begin{abstract}
Transverse  polarization of $\Lambda^0$ hyperons in inclusive
photoproduction at $x_F>0$ is tackled within the framework of the
quark recombination model, which has been successfully applied to
the polarization of different hyperons in a variety of unpolarized
hadron-hadron reactions. The results are compared with recent
experimental data of HERMES.
\end{abstract}
\bibliographystyle{apsrev}
\maketitle
\section{Introduction\label{introd}}

The problem of the $\Lambda^0$ polarization in hadron-hadron
reactions at high energies remains still vital even in spite of
the thirty years have passed since it was discovered
\cite{fermilab}. Being produced  in $pN$ collisions at 300 GeV
proton beam energy, the $\Lambda^0$ hyperons were found to be
highly polarized while neither the beam nor the beryllium target
possessed any initial polarization. Its direction was, in
accordance with the spatial parity conservation, opposite to the
unit vector $\vec{n}\propto[{\vec p_{b}}\times{\vec p_\Lambda}]$,
($\vec p_b$ and $\vec p_\Lambda$ are the beam and hyperon momenta,
respectively), which is normal to the production plane or, in
other words, transverse to the direction of this particle's
motion.

This phenomenon turned out to be quite surprising for the widely
spread belief that spin flip processes would not take any
significant place at such high energies as the helicity is
conserved in the limit of massless quarks.

Certainly, it has induced much attention to be focused as well on
studies of the polarization experimentally, using a variety of
beam hadrons and targets at different kinematic regimes, as on its
theoretical explanations. Thus, further experiments on $pN$
collisions in wide range of the beam energies were carried out
\cite{exp1, exp2,exp3,exp4,exp5,exp6}. The same was done for
$K^-p\rightarrow\Lambda X$ at $\sim$ 12 GeV - 176 GeV
\cite{exp17}, it was also examined in $\pi^-N\rightarrow\Lambda X$
and $K^+N\rightarrow\Lambda X$ \cite{exp19,exp20,exp21}. To obtain
more systematic knowledge on this issue, polarizations of other
hyperons were studied as well, e.g. $pN\rightarrow\Sigma^{0,\pm}
X$, $pN\rightarrow\Xi^{0,-}X$ \cite{exp7,
exp8,exp9,exp99,exp10,exp11,exp12,exp13,exp14,exp15,exp16}. A
different angle of sight, which could assist in the solution of
the problem, may be provided by processes, where hyperons
themselves acted as projectiles, for example
$\Sigma^-N\rightarrow\Lambda^0 X$,
$\Sigma^-N\rightarrow\Sigma^+X$, $\Sigma^-N\rightarrow\Xi^-X$
\cite{exp18}.

Among the most remarkable features of the $\Lambda^0$ polarization
one can highlight the extremely weak dependence on the incident
particle energy or, if the process is considered in the
center-of-mass (c.m.) frame, on the total c.m. energy $\sqrt{s}$.
The polarization grows by magnitude roughly linearly with the
transverse momentum of the hyperon $p_T$. It also depends, though
not so strongly as on $p_T$, on $x_F=2p_{L\Lambda}/\sqrt{s}$,
where $p_{z\Lambda}$ - longitudinal momentum of the outgoing
$\Lambda^0$. Another notable property is the sign of the
polarization, being negative in $pN$ collisions for $\Lambda^0$,
$\Xi^{0,-}$ it appears to be positive for $\Sigma^{0,\pm}$. The
positive sign has been observed in $K^-p\rightarrow\Lambda X$ as
well.

Although there have  been the large amount of experimental
information, no model is elaborated still to account convincingly
for the complete set of the available measurements from a unified
point of view. The existing phenomenological approaches are, in
more or less extent, fragmentary in reproducing the data (see,
e.g., Refs.
\cite{review,th1,degrand,swed,gago,th2,th22,th3,qrm1,qrm2,anselmino,th4}
and the references therein).

Especially useful instrument for spin effect investigations in
strong interactions seems to be the $\Lambda^0$ due to its wave
function structure peculiarities. The approximation of the SU(6)
symmetry requires the spin-flavor part of the wave function to be
combined of the $ud$ diquark in a singlet spin state and the
strange quark of spin 1/2, or rather formally
$|\Lambda\rangle_{1/2}=|ud\rangle_0|s\rangle_{1/2}$, where the
subscriptions refer to the spin states. Therefore, the $\Lambda^0$
total spin is entirely determined by its valence $s$ quark. Thus,
one may attribute the $\Lambda^0$ polarization to the strange
quark only \cite{swed, gago,th3}. It should be noted that the
SU(6) symmetric picture has been also applied to calculations of
the longitudinal $\Lambda^0$ polarization in $e^+e^-$ annihilation
at the $Z^0$ pole \cite{long1,long2} and then justified
experimentally \cite{long3,long4}.

In light of the discussion above, to wonder whether the
polarization would be manifested in reactions induced by pointlike
particles, such as leptons or photons, becomes an interesting
question. Indeed, experiments on high energy $\gamma N$ scattering
had been performed, for instance,  at CERN \cite{cern_gamma} and
SLAC \cite{slac_gamma} ($E_\gamma$=20 GeV - 70 GeV), however,
their statistical accuracy is insufficient for a decisive
conclusion on the magnitude or on the sign of the $\Lambda^0$
polarization. Rather relevant data for this purpose could be those
on the 27.6 GeV positron beam scattering from nucleon target
recently obtained by HERMES. The collaboration has measured
nonzero positive transverse $\Lambda^0$ polarization, herewith
most of the intermediate photons were very near the mass shell,
i.e. $Q^2=-(p_{ei}-p_{ef})^2\approx 0$ GeV$^2$, where
$p_{ei,f}$-are the 4-momenta of the initial and scattered
electrons, respectively (quasi-real photoproduction) \cite{Greb}.

We tackle here the transverse $\Lambda^0$ polarization in
inclusive photoproduction at $x_F>0$  in the framework of the
quark recombination model (QRM). Having been firstly proposed to
account for meson production probabilities in $pp$ collisions
\cite{das1,hwa}, the model was shown to be successful in
describing the polarizations of different hyperons in a variety of
high energy hadron-hadron reactions as well \cite{qrm1, qrm2}. We
discuss the quark recombination mechanism below.

\section{Quark Recombination Model \label{sec:model}}
%
\subsection{Key ingredients}
Let us, at first, briefly recall the essential ingredients of the
QRM concerning the hyperon polarization. One can find very
detailed description of the model in Ref. \cite{qrm1}. In the
sequel we will also abbreviate the collision $H_iN\rightarrow
H_fX$ (e.g. $K^-N\rightarrow\Lambda X$) as $H_i\rightarrow H_f$
($K^-\rightarrow\Lambda$).

The quantity proportional to the reaction probability of the
transition $H_i\rightarrow H_f$ in the projectile infinite
momentum frame (IMF) is defined as

\begin{multline}
|\langle M_f|S|M_i\rangle|^2\\=\sum_{s_k , \mu_k}
G^{M_f}_{4s_4\mu_4}(r_4) \otimes
G^{M_f}_{3s_3\mu_3}(r_3)\otimes|M(r_k;s_k,\mu_k)|^2\\ \otimes
G^{M_i}_{2s_2\mu_2}(r_2)\otimes
G^{M_i}_{1s_1\mu_1}(r_1)\otimes\Delta^3\otimes\Delta^4,
\label{eq:trans_prob}
\end{multline}

where $M_i$ and $M_f$ are the spin projections of the hadrons
$H_i$ and $H_f$ on the $z$ axis, which is defined by the vector
[$\vec p_{H_i}\times\vec p_{H_f}$], here $\vec p_{H_i}$ and $\vec
p_{H_f}$ are the momentum vectors of $H_i$ and $H_f$; the $x$ axis
is chosen to be parallel to $\vec p_{H_i}$; $r_k=(x_k,y_k,z_k)$
are the momentum fractions carried by the partons with respect to
the three independent directions $(x,y,z)$;
$G^{M_{i,f}}_{ks_k\mu_k}$ are the parton distribution functions,
the index $k$ denotes all the partons ($k$=1,2,3,4); the
summations are performed over the parton spins $s_k$ and their $z$
components $\mu_k$; $\Delta^3$ and $\Delta^4$ are the
delta-functions providing  energy-momentum conservation;
$|M(r_k;s_k,\mu_k)|^2$ is the squared amplitude of a parton-parton
scattering subprocess; the sign $\otimes$ denotes the convolution
in Bjorken $r_k$-space (see Eq. \ref{aeq:qrm4} in the appendix).

Then, the polarization is standardly given by

\begin{equation}
P=\dfrac{\sum\limits_{M_i}|\langle
+1/2|S|M_i\rangle|^2-\sum\limits_{M_i}|\langle
-1/2|S|M_i\rangle|^2}{\sum\limits_{M_i}|\langle
+1/2|S|M_i\rangle|^2+\sum\limits_{M_i}|\langle
-1/2|S|M_i\rangle|^2}.\label{eq:general}
\end{equation}

How the polarization will behave depends crucially on particular
forms of the squared amplitudes $|M|^2$, i.e. on the specification
of the underlying dynamic. Yamamoto, Kubo and Toki have calculated
them in Ref. \cite{qrm1} assuming a simple scalar type interaction
for the relativistic parton-parton scattering processes and noted
that non trivial spin dependent part appeared due to the
interference term between the lowest and higher order amplitudes,
similarly as in Refs. \cite{th3,feshb,dalitz}.

The final hadron of spin 1/2 may be resulted  in  recombinations
of a quark with a suitable diquark of spin 0 or of spin 1. Typical
representatives of such reactions, when considering the $x_F>0$
region, are the $p\rightarrow\Lambda$ ($(ud)_0$+$s$) and
$p\rightarrow\Xi^-$ ($d$+$(ss)_1$) transitions. Accordingly, there
are two free parameters in the model, $R_0$ - for scattering
between the partons of spin 1/2 and spin 0, $R_1$ - for scattering
between the partons of spin 1/2 and spin 1. Having been fixed to
fit the data for the transitions $p\rightarrow\Lambda$ and
$p\rightarrow\Xi^-$, the parameters were used to reproduce
reasonably the polarizations in other reactions of these kinds as
well, e.g. in $K^-\rightarrow\Lambda$ ($s$+$(ud)_0$),
$p\rightarrow\Xi^0$ ($u$+$(ss)_1$), $p\rightarrow\Sigma^+$
($(uu)_1$+$s$) and $\Sigma^-\rightarrow\Sigma^+$ ($s$+$(uu)_1$).

Another thing worthwhile to note is that the QRM automatically
contains the rule proposed by DeGrand and Miettinen
\cite{degrand}, and reproduces not only the magnitudes, but also
the signs of the polarizations.

\subsection{Applying to photoproduction}

\begin{figure}
\centering \centering
\includegraphics[width=0.4\textwidth]{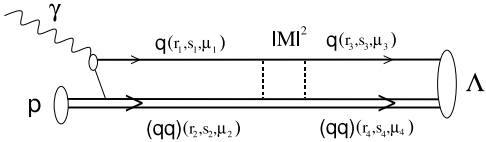}
\caption{Diagram corresponding to the transition
$\gamma\rightarrow q\bar q\rightarrow\Lambda$ in the QRM. To
produce the final $\Lambda^0$, a quark with quantum numbers
$(r_1,s_1,\mu_1)$ coming from the photon picks up an appropriate
diquark with the numbers $(r_2,s_2,\mu_2)$. The interaction is
entirely determined by the squared amplitude $|M|^2$.}
\label{Fig1}
\end{figure}

We turn now to the $\Lambda^0$ photoproduction at $x_F>0$. The QRM
can be straightforwardly extended to this process provided one
regards the photon as a hadron in the sense of its well known
quark degrees of freedom \cite{reya}. The corresponding diagram is
shown in Fig. \ref{Fig1}. To produce the final $\Lambda^0$, a
quark $q$ with the quantum numbers ($r_1$,$s_1$,$\mu_1$) coming
directly from the photon recombines with an appropriate diquark of
the proton with the numbers ($r_2$,$s_2$,$\mu_2$).

\begin{figure}
\centering \centering
\includegraphics[width=0.4\textwidth]{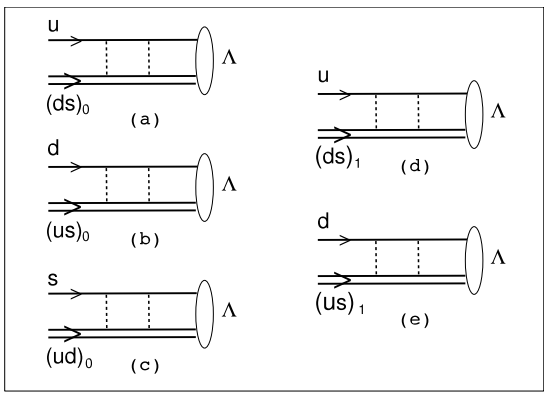}
\caption{Subprocesses of the $\Lambda^0$ photoproduction in the
QRM. One group of them, $(a)$, $(b)$ and $(c)$, concerns the
recombinations of quarks with scalar diquarks, $u$+$(ds)_0$,
$d$+$(us)_0$ and $s$+$(ud)_0$, respectively, while another one,
$(d)$ and $(e)$, refer to the recombinations of quarks with vector
diquarks, $u$+$(ds)_1$ and $d$+$(us)_1$. The subscriptions denote
the spin states.} \label{Fig2}
\end{figure}

Unlike a hadron-hadron reaction (say $p\rightarrow\Lambda$), which
is contributed, as a rule, by a single dominant subprocess
($(ud)_0$+$s$), the situation for the $\gamma\rightarrow\Lambda^0$
transition can be fairly expected to be rather rich. The most
probable scenarios we have assumed for this case are presented in
Fig. \ref{Fig2}, the pictures $(a)$, $(b)$ and $(c)$ concern the
recombinations of quarks with scalar diquarks (scalar case),
$u$+$(ds)_0$, $d$+$(us)_0$ and $s$+$(ud)_0$, respectively, while
the $(d)$ and $(e)$ refer to the recombinations of quarks with
vector diquarks, $u$+$(ds)_1$ and $d$+$(us)_1$ (vector case).

%

Applying of Eqs. (\ref{eq:trans_prob}) and (\ref{eq:general}) to
the $\Lambda^0$ photoproduction leads to the following formula for
the polarization \cite{qrm2}

\begin{equation}
P=\dfrac{\sum\limits_{i,j,k}\sum\limits_{l}R_lJ_{D}^{lijk}}
{\sum\limits_{i,j,k}\sum\limits_{l}J_{I}^{lijk}},\label{eq:qrm3}
\end{equation}

where $R_l$ are the free parameters, so that the corresponding sum
is performed over the scalar ($l=0$) and vector ($l=1$) cases,

\begin{multline}
J_{D(I)}^{lijk}=G_{\Lambda}^2\otimes\sigma_{D(I)}^l\otimes
f^p_{(q_iq_j)_l}\otimes
f^{\gamma}_{q_k}\otimes\Delta^3\otimes\Delta^4.\label{eq:qrm4}
\end{multline}

Here, $G_{\Lambda}$ is the light cone wave function of $\Lambda^0$
\cite{lapage}; $\sigma_{D}^l$ is the interference term surviving
in the numerator of Eq. (\ref{eq:general}); $\sigma_{I}^l$ is the
quantity proportional to the total probability in the denominator
of the same equation; $f^p_{(q_iq_j)_l}$ is the momentum
distribution function of the $(q_iq_j)_l$ diquark in the proton;
$f^{\gamma}_{q_k}$ is the structure function of the photon. The
sum over $i,j,k$ is rather symbolic and includes only the
appropriate combinations of quarks and diquarks  to form the final
$\Lambda^0$ (see Fig. \ref{Fig2}).

Note that, in the QRM, the distribution functions are factorized
into longitudinal and transverse momentum distribution parts as

\begin{equation}
f(r_k)=f(x_k,y_k,z_k)=f(x_k)e^{-(y_k^2+z_k^2)}.\label{eq:factorize}
\end{equation}

Having taken the transverse parts of all the functions to have the
same Gaussian form, we discuss henceforth the longitudinal those.

\section{Calculations and results}

We present here the results of the QRM calculations of the
$\Lambda^0$ polarization in photoproduction at $x_F>0$.

We used the Eqs. (\ref{eq:qrm3})-(\ref{eq:qrm4}). Explicit
expressions for $\sigma_{D(I)}^l$ as well as the parameter values
were taken from Ref. \cite{qrm1}. It should be emphasized that all
the parameters have been fixed for consistent fitting of the
polarization in a variety of hadron-hadron reactions. Thus, for
the $u$+$(ds)_0$, $d$+$(us)_0$ and $s$+$(ud)_0$ cases we took
$R_0=2.5$ GeV, and for the $u$+$(ds)_1$, $d$+$(us)_1$ it was
$R_1=-5.6$ GeV.

\begin{figure}
\centering \centering
\includegraphics[width=0.4\textwidth]{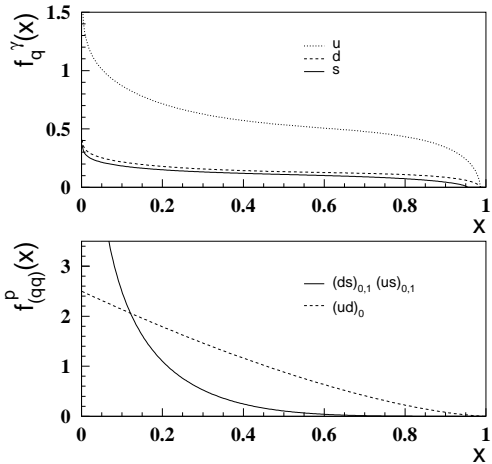}
\caption{Upper panel: The photon structure function. The
probabilities to find $u$ quark (dotted line), $d$ quark (dashed
line) and $s$ quark (solid line) in the photon at $Q^2=8$ GeV$^2$
\cite{reya}. Lower panel: Diquark distribution functions of the
proton. The probabilities to find $(ds)_{0,1}$ and $(us)_{0,1}$ in
the proton are assumed to be the same (solid line) except for
$(ud)_0$ (dashed line) \cite{ekelin}.} \label{Fig3}
\end{figure}

The photon structure function $f^{\gamma}_{q_k}$ plotted in the
upper panel of Fig. \ref{Fig3} is taken from Ref. \cite{reya}, the
probabilities to find $u$, $d$ and $s$ quarks in the photon (up to
a factor which does not affect the results since we deal with the
ratio (\ref{eq:qrm3})) are given by the dotted, dashed and solid
lines, respectively. For the diquark distribution functions of the
proton $f_{(q_iq_j)_l}^p$ we adopted those from Ref. \cite{ekelin}
shown in the lower panel of Fig. \ref{Fig3}. We assumed that the
functions for scalar and vector diquarks coincide (solid line)
except for $(ud)_0$ (dashed line) due to the valence character of
both $u$ and $d$ quarks forming it. Note that the functions depend
on the momentum transfer squared and we have taken them at $Q^2=8$
GeV$^2$.

We have chosen the masses of quarks to be the following
$m_u=m_d=0.3$ GeV, $m_s=0.55$ GeV, those of diquarks being simply
the sums of the corresponding quark masses, i.e.
$m_{(us)_{0,1}}=m_{(ds)_{0,1}}=m_u+m_s=0.85$ GeV and
$m_{(ud)_0}=0.6$ GeV. Other fixed quantities of the QRM  are the
confinement scale parameter $\beta$=0.5 GeV in the $\Lambda^0$
light cone wave function and a parameter $p_t=0.3$ GeV, which
fixed the transverse momentum distribution of the partons.

\begin{figure}
\centering \centering
\includegraphics[width=0.4\textwidth]{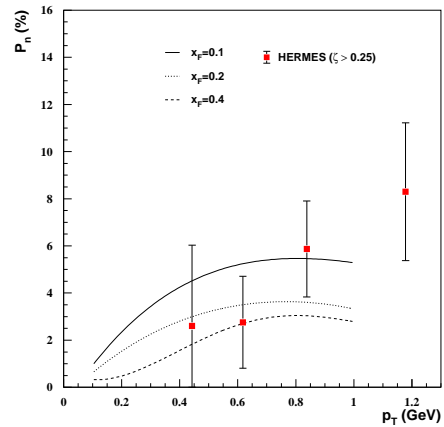}
\caption{The calculated dependence of the $\Lambda^0$ polarization
on $p_T$ at $x_F=0.1$ (solid line), $x_F=0.2$ (dotted line) and
$x_F=0.4$ (dashed line) in comparison with the HERMES data from
Ref. \cite{Greb} (solid points). } \label{Fig4}
\end{figure}

The calculated $p_T$ dependence of the polarization in the range
0.1 GeV $\leq p_T\leq$ 1.0 GeV is shown in Fig. \ref{Fig4} at
$x_F=0.1$ (solid line), $x_F=0.2$ (dotted line) and $x_F=0.4$
(dashed line). First of all, one can see the polarization turn out
to be positive. It grows more rapidly at lower $p_T$'s reaching
approximate plateaus at about $p_T=0.6$ GeV, which is more
distinctly manifested at $x_F=0.1$. The polarization decreases as
one considers the higher $x_F$'s. It is seen that the calculations
are in a good agreement with the HERMES data at $\zeta>0.25$
(solid points). The definition of the variable $\zeta$ will be
given later.

\begin{figure}
\centering \centering
\includegraphics[width=0.4\textwidth]{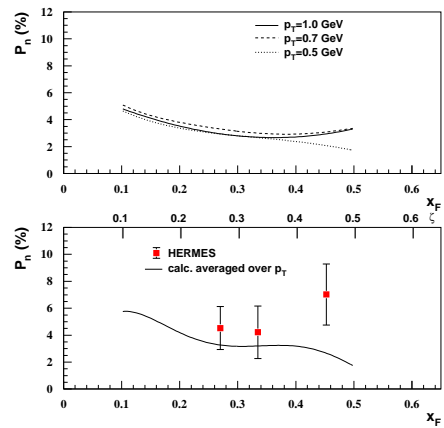}
\caption{Upper panel: The calculated dependence of the $\Lambda^0$
polarization on $x_F$ at $p_T=0.5$ GeV (dotted line), $p_T=0.7$
GeV (dashed line) and $p_T=1.0$ GeV (solid line). Lower panel: The
calculated dependence of the $\Lambda^0$ polarization on $x_F$
averaged over the $p_T$ distribution of $\Lambda^0$'s produced at
HERMES (solid line) in comparison with the experimental dependence
on $\zeta$ (solid points). The data are taken from Ref.
\cite{Greb}.} \label{Fig5}
\end{figure}

The calculated $x_F$ dependence in the range $0.1\leq x_F\leq0.5$
is presented in the upper panel of Fig. \ref{Fig5} at $p_T=0.5$
GeV (dotted line), $p_T=0.7$ GeV (dashed line) and $p_T=1.0$ GeV
(solid line). One can see that the lines corresponding to the
three values of $p_T$  fall slowly as $x_F$ increases up to about
$x_F=0.34$, being, herewith, very close one to other. Afterwards,
the line concerning $p_T=0.5$ GeV branches off the common trend
and continues to fall while the rest those begin weakly to rise.

In Fig. \ref{Fig4} we have demonstrated how these calculations
related to the HERMES measurements on the $\Lambda^0$ polarization
in quasi-real photoproduction, which seem to be more suitable for
this purpose \cite{Greb}. However, we should make at this point a
few comments. For some peculiarities of the HERMES experiment, the
data are collected not as the traditional $x_F$ dependence but as
the dependence on
$\zeta=(E_{\Lambda}+p_{L\Lambda})/(E_{b}+p_{Lb})$, additionally
integrated over $p_T$ ($E_b$ and $p_{Lb}$ are the energy and
longitudinal momentum of the beam particle). Unlike $x_F$, the
variable $\zeta$ is, thus, just an approximate measure of whether
the hyperons were produced in the current or target fragmentation
regions. Hence there is some ambiguity in the correlation between
$x_F$ and $\zeta$, which causes an arbitrariness in the comparison
of the HERMES data with results expressed in terms of $x_F$. The
experimental $p_T$ dependence is also collected integrally over
$\zeta$ for two regions, $\zeta\leq0.25$ and $\zeta>0.25$.
Additionally, the intermediate quasi-real photons of HERMES were
not, certainly, monoenergetic, though this problem could be
omitted by exploiting the fact that the polarization is incident
particle energy independent.

To make the comparison with the experiment more correct, we have
averaged the calculated $x_F$ dependence of the polarization over
the $p_T$ distribution of $\Lambda^0$ hyperons produced at HERMES
\cite{ptdistr}. We show thus obtained results in the lower panel
of Fig. \ref{Fig5} (solid line) in comparison with the
experimental $\zeta$ dependence of the $\Lambda^0$ polarization
(solid points). We used  only the HERMES events at $\zeta>0.25$
because they  more adequately relate to the $x_F>0$ region. One
can see that the calculations sufficiently reproduce the
experimental events.

%
\section{Summary and discussion \label{conclusions}}

Following the recipes given in Refs. \cite{qrm1,qrm2}, we have
shown that the transverse $\Lambda^0$ polarization in inclusive
photoproduction at $x_F>0$ can be fairly accommodated by the quark
recombination model, which comes, thus, outside of the reactions
induced by hadrons. All the free parameters we used in the
calculations have been already fixed to reproduce the polarization
in other hadron-hadron interactions.

We have calculated as well the $p_T$ dependence of the
polarization at $x_F=0.1$, $x_F=0.2$ and $x_F=0.4$ in the range
0.1 GeV $\leq p_T\leq$ 1.0 GeV as the dependence on $x_F$ at three
fixed values of $p_T$, $p_T=0.5$ GeV, $p_T=0.7$ GeV, $p_T=1.0$ GeV
in the range 0.1 $\leq x_F\leq$ 0.5.

We have compared the results with the HERMES data and discussed in
what extent it could be suitable for this purpose. It was stressed
that there is some ambiguity between the data and the results
expressed in terms of $x_F$. To obtain results, which could be
more correctly comparable with the experiment, we have averaged
the calculated $x_F$ dependence over the $p_T$ distribution of
$\Lambda^0$ hyperons produced at HERMES. Additionally, we used
only the events of $\zeta>0.25$ to be, more or less, ensured that
we dealt with the region of $x_F>0$.  So, we have found a
sufficient agreement with the data both in magnitude and in the
sign of the polarization. However, this consistency can be
regarded only as qualitative because of, at least, a few reasons.
The uncertainties associated with the correlations between $\zeta$
and $x_F$ still remain. The intermediate photons emitted by the
HERMES positron beam were not, indeed, monoenergetic. No
information on the momentum transfer squared  was derivable at the
experiment while the structure functions used here are $Q^2$
dependent.

There are also another problems. Since the spin dependent
distributions of the partons were not available, we have naively
assumed the structure functions to have the same form as well for
the scalar as for vector diquarks. We have also supposed that the
subprocesses $s$+$(ud)_0$, $d$+$(us)_0$ and $u$+$(ds)_0$
contributed in the polarization with the same, positive, sign.
These cases are structurally similar to the
$K^-\rightarrow\Lambda^0$, $\pi^-\rightarrow\Lambda^0$ and
$K^+\rightarrow\Lambda^0$ transitions, respectively. Certainly,
the positive sign has been very reliably determined for
$K^-\rightarrow\Lambda^0$, while, in fact, for the rest two cases
the related situation is controversial due to the error bars of
the data are still large (see also discussion in Ref. \cite{qrm1}
). In this light, it would be interesting to compare our results
with those from Ref. \cite{suzuki}, where similar calculations
have been carried out.

We did not take here contributions from the heavier resonances
into account, which are presumably significant for the $\Lambda^0$
polarization \cite{gatto,long1,th4,liang}. It can be done as a
further improvement of the calculations, but, for this purpose,
one needs to know, at least, the evolution of the variables $x_F$
and $p_T$ in the transition processes from the resonances to the
final $\Lambda^0$.

It seems to be attractive to find the explicit expressions of the
QRM amplitudes specifying the potential by the color field
\cite{th3}, which might lead, in some sense, to a unification of
the present approach with other quark scattering models
\cite{swed,gago}.

We would like to thank K. Suzuki for providing useful information
on the quark recombination model.

\appendix*
\section{}

We present here some steps of the calculations in more explicit
form.

The convolution in Eq. (\ref{eq:qrm4}) is defined by

\begin{multline}
J_{D(I)}^{lijk}=G_{\Lambda}^2\otimes\sigma_{D(I)}^l\otimes
f^p_{(q_iq_j)_l}\otimes
f^{\gamma}_{q_k}\otimes\Delta^3\otimes\Delta^4\\=\int\left[\prod\limits_
{m=1}^4\dfrac{dx_mdy_mdz_m}{x_m}\right]G_{\Lambda}^2\,\sigma_{D(I)}^l\,f^p_{(q_iq_j)_l}\,
f^{\gamma}_{q_k}\,\Delta^3\,\Delta^4,\label{aeq:qrm4}
\end{multline}

so that the integral is 12-dimensional.

According to Ref. \cite{qrm1}, we take

\begin{multline}
\Delta^3=
\delta(x_Fx_4+x_Fx_3-x_F)\\\times\delta(y_4+y_3-P_T/p_t)\delta(z_4+z_3),
\label{delta3}
\end{multline}

\begin{multline}
\Delta^4=\delta(x_Fx_3+x_Fx_4-x_1-x_2)\delta(y_3+y_4-y_1-y_2)\\\times
\delta(z_3+z_4-z_1-z_2)\delta\left(E_{Tf}-E_{Ti}\right),\label{delta4}
\end{multline}

where $P_T$ is the transverse momentum of $\Lambda^0$, $p_t$ is a
normalization parameter to fix the transverse momentum
distribution of the partons,

$$E_{Tf}=\dfrac{(y^2_3+z^2_3)p^2_t+m^2_{q}}{x_Fx_3}+
\dfrac{(y^2_4+z^2_4)p^2_t+m^2_{(qq)}}{x_Fx_4},$$
$$E_{Ti}=\dfrac{(y^2_1+z^2_1)p^2_t+m^2_{q}}{x_1}+
\dfrac{(y^2_2+z^2_2)p^2_t+m^2_{(qq)}}{x_2}.$$

We realized the condition when all the hyperons would be produced
in the $x_F>0$ region by formal introducing the step function
$\theta(x_1-x_2)$, which simply means that each quark coming from
the photon will be faster than the corresponding picked up
diquark.

Let us rewrite Eq. (\ref{aeq:qrm4}) as

\begin{multline}
J=\int\left[\prod\limits _{m=1}^4{dx_mdy_mdz_m}\right]
F(r_1,r_2,r_3,r_4)\Delta^3\Delta^4,\label{aeq:qrm5}
\end{multline}

where

\begin{equation}
F(r_1,r_2,r_3,r_4)=\dfrac{G_{\Lambda}^2\,\sigma_{D(I)}^l\,f^p_{(q_iq_j)_l}\,
f^{\gamma}_{q_k}}{x_1x_2x_3x_4}\,\theta(x_1-x_2).\label{denot}
\end{equation}

To concentrate the attention on the integration over the momentum
fractions $r_m=(x_m,y_m,z_m)$, we introduced the denotation
(\ref{denot}). For the same reason, the dependences on the rest
parameters and indices are omitted.

We reduced the 12-dimensional integral to 5-dimensional one by
using the delta functions (\ref{delta3}) and (\ref{delta4}).

Thus, an integration over $x_1, x_2, y_1, y_3, z_1, z_3$ leads to
the following substitutions in Eq. (\ref{denot})

\begin{equation}
h=\left\{\begin{aligned}x_1=x_F-x_2\\
x_3=1-x_4\\
y_1=\dfrac{P_T}{p_t}-y_2\\
y_3=\dfrac{P_T}{p_t}-y_4\\
z_1=-z_2\\
z_3=-z_4.
\end{aligned}\right.\label{eq:apcond}
\end{equation}

Using the remaining delta-function
$\delta\left(E_{Tf}-E_{Ti}\right)$, we integrated over $z_4$ as
follows

\begin{equation}
J=\left.\dfrac{1}{x_F}\int\dots dz_4
F(\dots,z_4)\right|_{h}\delta(az_4^2-b),\label{aeq:qrm5}
\end{equation}

where $h$ denotes the conditions (\ref{eq:apcond}),

\begin{equation}
a=\frac{p_t^2}{x_Fx_4(1-x_4)}, \label{a}
\end{equation}

\begin{multline}
b=\frac{((\frac{P_T}{p_t}-y_2)^2+z^2_2)p^2_t+m^2_{q}}{x_F-x_2}+
\frac{(y^2_2+z^2_2)p^2_t+m^2_{(qq)}}{x_2}\\-\frac{(\frac{P_T}{p_t}-y_4)^2p^2_t+m^2_{q}}{x_F(1-x_4)}
-\frac{y_4^2p^2_t+m^2_{(qq)}}{x_Fx_4}. \label{b}
\end{multline}

Applying the well known property of the delta-function one can
write that

\begin{equation}
\delta(az_4^2-b)=\frac{1}{2}\sqrt{\frac{1}{ab}}\left[
\delta\left(z_4-\sqrt{\frac{b}{a}}\right)+\delta\left(z_4+\sqrt{\frac{b}{a}}\right)\right].
\end{equation}

Finally, after the integration over $z_4$, the Eq.
(\ref{aeq:qrm5}) is split into a sum of integrals to be calculated
numerically,

\begin{equation}
J=J^{+}+J^{-},
\end{equation}

where

\begin{multline}
J^{\pm}=\dfrac{1}{2x_F}
\int\limits_{\varepsilon}^{\frac{x_F}{2}}dx_2\int\limits_{\varepsilon}^{1-\varepsilon}dx_4
\int\limits_{-\infty}^{\infty}dy_2\int\limits_{-\infty}^{\infty}dy_4\int\limits_{-\infty}^{\infty}dz_2
\sqrt{\dfrac{1}{ab}}
\\\left.\frac{}{}\times
F(r_1,r_2,r_3,r_4)\right|_{\stackrel{h}{z_4}=\pm\sqrt{\frac{b}{a}}}.
\label{eq:appm}
\end{multline}

The integration limit $\dfrac{x_F}{2}$ arose due to the
step-function in Eq. (\ref{denot}), $\varepsilon$ is introduced
because of the difficulties associated with the irregular behavior
of the integrand at the borders of the integration regions over
$x_2$ and $x_4$. In the actual computations, we have taken
$\varepsilon=0.01$.


%
%

\begin{thebibliography}{99}
%
\bibitem{fermilab}
G. Bunce \textit{et al}., Phys. Rev. Lett. {\bfseries 36}, 1113
(1976).
%
\bibitem{exp1} K. Heller \textit{et al}., Phys. Lett. B {\bfseries 68}, 480 (1977).
%
\bibitem{exp2} L. G. Pondorm, Phys. Rep. {\bfseries 122}, 57 (1985).
%
\bibitem{exp3} B. Lundberg \textit{et al}., Phys. Rev. D {\bfseries 40}, 3557 (1989).
%
\bibitem{exp4} S. Erhan \textit{et al}., Phys. Lett. B {\bfseries 82}, 301 (1979).
%
\bibitem{exp5} F. Abe \textit{et al}., Phys. Rev. Lett. {\bfseries 50}, 1102 (1983).
%
\bibitem{exp6} A. M. Smith \textit{et al}., Phys. Lett. B {\bfseries 185}, 209 (1987).
%
\bibitem{exp17} S. A. Gourly \textit{et al}., Phys. Rev. Lett. {\bfseries 56}, 2244 (1986).
%
\bibitem{exp20} I. V. Ajinenko \textit{et al}., Phys. Lett. B {\bfseries 121}, 183 (1983).
%
\bibitem{exp19} M. L. Faccini-Turluer \textit{et al}., Z. Phys. C {\bfseries 1}, 19 (1979).
%
\bibitem{exp21} J. Bensinger \textit{et al}., Phys. Rev. Lett. {\bfseries 50}, 313 (1983).
%
\bibitem{exp7} K. Heller \textit{et al}., Phys. Rev. Lett. {\bfseries 51}, 2025 (1983).
%
\bibitem{exp8} R. Rameika \textit{et al}., Phys. Rev. D {\bfseries 33}, 3172 (1986).
%
\bibitem{exp9} L. H. Trost \textit{et al}., Phys. Rev. D {\bfseries 40}, 1703 (1989).
%
\bibitem{exp99} J. Duryea \textit{et al}., Phys. Rev. Lett. {\bfseries 67}, 1193 (1991).
%
\bibitem{exp10} C. Wilkinson \textit{et al}., Phys. Rev. Lett. {\bfseries 46}, 803 (1981).
%
\bibitem{exp11} C. Ankenbrandt \textit{et al}., Phys. Rev. Lett. {\bfseries 51}, 863 (1983).
%
\bibitem{exp12} C. Wilkinson \textit{et al}., Phys. Rev. Lett. {\bfseries 58}, 855 (1987).
%
\bibitem{exp13} A. Morelos \textit{et al}., Phys. Rev. Lett. {\bfseries 71}, 2172 (1993).
%
\bibitem{exp14} E. C. Dukes \textit{et al}., Phys. Lett. B {\bfseries 193}, 135 (1987).
%
\bibitem{exp15} L. Deck \textit{et al}., Phys. Rev. D {\bfseries 28}, 1 (1983).
%
\bibitem{exp16} Y. W. Wah \textit{et al}., Phys. Rev. Lett. {\bfseries 55}, 2551 (1985).
%
\bibitem{exp18} M. I. Adamovich \textit{et al}., Z. Phys. A {\bfseries 350}, 379 (1995).
%
\bibitem{review} for reviews of experimental status and existing models, see

A. D. Panagiotou, Int. J. Mod. Phys. {\bfseries A5}, 1197 (1990).

J. Soffer, hep-ph/9911373.

S. M. Troshin, N. E. Tyurin, hep-ph/0201267

%
\bibitem{th1} B. Andersson, G. Gustafson and G. Ingelman, Phys. Lett. B {\bfseries 85},
417 (1979).
%
\bibitem{degrand} T. A. DeGrand and H. I. Miettinen, Phys. Rev. D {\bfseries 23}, 1227 (1981).
%
\bibitem{swed} J. Szwed, Phys. Lett. B {\bfseries 105}, 403 (1981).
%
\bibitem{gago} J. M. Gago R. V. Mendes and P. Vaz, Phys. Lett. B {\bfseries 183},
 357 (1987).
%
\bibitem{th2} S. Soffer and N. A. T\"orngvist, Phys. Rev. Lett. {\bfseries 68},
 907 (1992).
%
\bibitem{th22} C. Boros, Liang Zuo-tang and Meng Ta-chung, Phys. Rev. Lett. {\bfseries 70},
1751 (1993).
%
\bibitem{th3}  W. G. D. Dharmaratna and G. R. Goldstein, Phys. Rev. D {\bfseries 53},
1073 (1996).
%
\bibitem{qrm1} Y. Yamamoto, K.-I. Kubo and H. Toki, Prog. Theor. Phys. {\bfseries 98}, 95
(1997).
%
\bibitem{qrm2} N. Nakajima, K. Suzuki, H. Toki and K.-I. Kubo, hep-ph/9906451.
%
\bibitem{anselmino}  M. Anselmino, D. Boer, U. D'Alesio and F. Murgia, Phys. Rev. D {\bfseries 63},
054029 (2001).
%
\bibitem{th4} Dong Hui and Liang Zuo-tang, Phys. Rev. D {\bfseries 70}, 014019
(2004).
%
\bibitem{long1} G. Gustafson and J. H$\ddot{a}$kkinen,  Phys. Lett. B {\bfseries 303}, 350 (1993).
%
\bibitem{long2} C. Boros and Liang Zuo-tang, Phys. Rev. D {\bfseries 57},
4491 (1998).
%
\bibitem{long3} ALEPH Collaboration, D. Buskulic \textit{et al}.,  Phys. Lett. B {\bfseries 374}, 319 (1996).
%
\bibitem{long4} OPAL Collaboration, K. Ackerstaff \textit{et al}., Eur. Phys. J. C {\bfseries 2}, 49 (1998).
%
\bibitem{cern_gamma}
CERN-WA-004 Collaboration, D. Aston \textit{et al}., Nucl. Phys.
{\bfseries B195}, 189 (1982).
%
\bibitem{slac_gamma}
SLAC-BC-072 Collaboration, K. Abe \textit{et al}., Phys. Rev. D
 {\bfseries 29}, 1877 (1984).
%
\bibitem{Greb} HERMES Collaboration, A. Airapetian \textit{et al}., arXiv:0704.3133v1
[hep-ex].
%
\bibitem{das1} K. P. Das and R. C. Hwa,  Phys. Lett. B {\bfseries 68}, 459 (1977).
%
\bibitem{hwa} R. C. Hwa,  Phys. Rev. D {\bfseries 22}, 1593 (1980).
%
%
\bibitem{reya} M. Gl\"{u}ck, E. Reya and A. Vogt, Phys. Rev. D {\bfseries 45}, 3986 (1992).
%
\bibitem{lapage} G. P. Lepage and S. J. Brodsky, Phys. Rev. D {\bfseries 22}, 2157 (1980).
%
\bibitem{ekelin} S. Ekelin and S. Fredrikson, Phys. Lett. B {\bfseries 162}, 373 (1985).
%
\bibitem{feshb} A. W. McKinley and H. Feshbach, Phys. Rev. {\bfseries 74},   1759 (1948).
%
\bibitem{dalitz} R. H.  Dalitz, Proc. R. Soc. (London) {\bfseries A206},  509 (1951).
%
\bibitem{suzuki} K.-I. Kubo, K. Suzuki, hep-ph/0505179.
%
\bibitem{ptdistr}
HERMES Collaboration, S. Belostotski, O. Grebenyuk and Yu.
Naryshkin, Acta Phys. Polon. B {\bfseries 33}, 3785 (2002).
%
\bibitem{gatto} R. Gatto, Phys. Rev. {\bfseries 109}, 610 (1958).
%
\bibitem{liang} Liang Zuo-tang and Liu Chun-xiu, Phys. Rev. D {\bfseries 66}, 057302
(2002).
%
\end{thebibliography}
\end{document}